# Topological transport of sound mediated by spin-redirection geometric phase

Short title: **Acoustic spin-redirection phase**

Shubo Wang[1,2], Guancong Ma[1], and Che Ting Chan[1]*

[1]Department of Physics, The Hong Kong University of Science and Technology, Hong Kong, China.

[2]Department of Physics, City University of Hong Kong, Hong Kong, China.

* Correspondence to C. T. Chan (phchan@ust.hk)

When a dynamic system undergoes a cyclic evolution, a geometric phase that depends only on the path traversed in parameter space can arise in addition to the normal dynamical phase. These geometric phases have profound impacts in both quantum and classical physics. In addition to the geometric phase associated with band structures in reciprocal space that has led to the discovery of topological insulators, the spin-redirection geometric phase induced by the *SO*(3) rotation of states in real space can also give rise to intriguing phenomena such as the photonic analog of spin Hall effect. By exploiting the orbital angular momentum of sound vortices, we theoretically and experimentally demonstrate the spin-redirection geometric phase effects in airborne sound, which is a scalar wave without spin. We show that these effects, associated with the helical transport of sound, can be used to control the flow of sound. This finding opens new possibilities for the manipulation of scalar wave propagation by exploiting spin-redirection geometric phases.

## INTRODUCTION

Is there a simple way to know whether a buried pipe is twisted? If the same question is posed for an optical fiber, the answer is well known. One can send circularly polarized lights down the fiber and the output phase will be different for right and left circular polarizations if the fiber is twisted in space (*1*). The difference is attributed to the geometric phase. To understand such a phase, consider a Hamiltonian $H(\lambda)$, which smoothly depends on the parameter $\lambda$ as in $H(\lambda)\Psi = E(\lambda)\Psi$, where $E(\lambda)$ is the nondegenerate eigenvalue. For the adiabatic evolution of the Hamiltonian with respect to the parameter $\lambda$, the variation of $\lambda$ can induce a coupling between a fast and a slow variable of the system. By separately treating the two variables as in the Born-Oppenheimer approximation, the dynamics of the fast variable can induce an effective vector potential (the curl of which defines an effective magnetic field) acting on the slow dynamic motion of the system (*2*). Such a vector potential contributes to a phase factor that corresponds to the $U(1)$ Abelian gauge field and is essentially the holonomy that characterizes the twisting of a Hermitian line bundle over parameter space (*3*), which is the geometric phase. In the case of degenerate eigenvalues, the adiabatic evolution can give rise to a non-Abelian gauge structure (*4*). The concept of geometric phase applies to both quantum-mechanical (*5*, *6*) and classical systems (*7–18*) and has revolutionized our understanding of state evolutions in different parameter spaces. In classical systems, the geometric phase associated with band structures in reciprocal space has led to the discovery of photonic/acoustic topological insulators (*19–21*). In contrast, geometric phase can also arise from $SO(3)$ rotation of states in real space, which is called spin-redirection geometric phase and it gives rise to intriguing phenomena such as the photonic analog of spin/orbital Hall effect (*22–25*).

Here, we show that the spin-redirection geometric phase effect can also be realized with airborne sound. It may sound paradoxical because sound propagates in fluids as a longitudinal scalar wave and the absence of a polarization vector means that sound cannot give rise to geometric phases through the spin degree of freedom. However, a sound wave can carry orbital angular momentum (OAM) in the form of a sound vortex (*26–29*), which provides a new degree of freedom for the manifestation of the spin-redirection geometric phase effect. We show that the transport of sound vortices inside a helical waveguide induces geometric phases that can affect the transmission properties. These phenomena can be viewed as a consequence of the coupling between the OAM and the linear momentum induced by helical transport.

**RESULTS**

**Spin-redirection geometric phase of sound**

We consider a circular waveguide of radius $a$ filled with air. Sound propagating inside the waveguide is governed by the wave equation $c^2 \nabla^2 P(\mathbf{r},t) = \partial^2 P(\mathbf{r},t) / \partial t^2$, where $c$ is the sound speed and $P(\mathbf{r},t)$ is the pressure field. By defining the operator $\hat{H} = c^2 \nabla^2$, the wave equation can be rewritten as $\hat{H} P(\mathbf{r},t) = \partial^2 P(\mathbf{r},t) / \partial t^2$. Using cylindrical coordinates $(r, \varphi, z)$ and the separation of variables, one can obtain the solutions to the wave equation for the cylindrical waveguide in the form of $J_l(k_r r)(A\cos l\varphi + B \sin l\varphi)e^{ik_z z}e^{i\omega t}$ (*30*), where $J_l(k_r r)$ is the $l$th-order cylindrical Bessel function of the first kind; $k_r$ and $k_z$ are the transverse and longitudinal wavenumber, respectively; and $A$ and $B$ are arbitrary coefficients. Because the radial particle velocity at $r=a$ is zero, the Dirichlet boundary condition $[\partial J_l(k_r r)/\partial r]_{r=a} = 0$ is applied to determine the discrete guided modes. These modes can be labeled by $(m,l)$ with $m$ denoting the $m^{th}$ root. Figure 1 shows the dispersion relations of the three lowest-order modes with the corresponding mode pressure field. The monopole mode has a homogeneous pressure distribution along the transverse plane and, hence, the pressure variation is purely longitudinal. The pressure fields of higher-order modes, such as the dipole mode, have nonuniform phases on the transverse plane and therefore can be assigned a transverse polarization, so that the guided sound modes can be viewed as

"transversely" polarized (*30*). We note that the solutions to the wave equation can be expanded using the complete basis $\{J_l(k_r r)e^{il\varphi}e^{ik_z z}\}$. For the partial waves with $m=0$ and $l\neq 0$, $J_l(k_r r)e^{il\varphi}e^{ik_z z}$ denotes a vortex field of topological charge $q=l$ that carries an OAM of $l\hbar$ per quantum. A sound vortex with $q=+1$ can be excited using four monopole sources with an azimuthal phase gradient of $\pi/2$ at one end of the waveguide, as shown in the inset of Fig. 1, where the excited pressure field is plotted. Such a sound vortex can also be excited using a transducer array (*26, 31*), a metasurface (*28, 29*) or resonance structures (*27*). A sound vortex with an opposite charge of $q=-1$ can be excited by simply reversing the phase gradient of the monopole sources.

Now, we bend the straight waveguide into a helix of radius $R$ and pitch $D$, as shown in Fig. 2A. The helical structures have been widely used in electromagnetics (such as antennae (*32*) and traveling-wave tube (*33*)) and in photonics and plasmonics (*34–36*). However, the helical transport of sound remains largely unexplored. The helical bending effectively induces a rotation of the local coordinate frame $(\mathbf{u},\mathbf{v},\mathbf{t})$ attached to the waveguide, where $\mathbf{t}$ denotes the unit tangent vector that forms an angle $\theta$ with the laboratory $z$ axis. Therefore, under adiabatic conditions, the wave undergoes the transformation $P'=\hat{R}(\mathbf{n},\phi)P$, where $\hat{R}(\mathbf{n},\phi)=e^{-i\phi\hat{L}\cdot\mathbf{n}}$ is the unitary rotation operator that applies a rotation of angle $\phi$ on $P$ with respect to the axis $\mathbf{n}$. $\hat{L}=-i\mathbf{r}\times\nabla$ is the OAM operator that serves as the generator of the rotation. The time evolution of the vortices can be expressed as (see Materials and Methods)

$$\frac{\partial^2 P'}{\partial t^2}=\left[\hat{H}+2\omega\hat{L}\cdot\mathbf{\Omega}-\left(\hat{L}\cdot\mathbf{\Omega}\right)^2\right]P', \qquad (1)$$

where in the laboratory frame we have $\hat{L}\cdot\mathbf{\Omega}=\mathbf{t}\cdot\hat{L}\Omega(\cos\theta-1)$ and $\Omega=\partial\phi/\partial t$. Using the relationship $\hat{p}=-i\nabla$, the operator for the helical waveguide can be expressed formally as

$$\hat{H}'(\mathbf{k})=-c^2\left(\hat{p}-\alpha\hat{L}\cdot\mathbf{k}\right)^2, \qquad (2)$$

where $\alpha = |\mathbf{\alpha}| = (\cos\theta - 1)\sin\theta c/(R\omega)$ and $\mathbf{\alpha}$ is in the same direction as $\hat{p}$; and $\mathbf{k} = k\mathbf{t}$ is the longitudinal wave vector inside the waveguide. The corresponding eigen frequency is

$$\omega' = \omega - c\alpha l k. \tag{3}$$

If we view the helical waveguide as a periodic structure, the longitudinal wave vector $\mathbf{k}$ in Eq. (2) can be mapped to the Bloch wave vector $\mathbf{k}_B = k_B \hat{z}$ with $k = (D/S)k_B$, where $S$ is the total length of the helical waveguide for one pitch (see Fig. S1 and related discussions in Supplementary Materials). In the limit of $a \ll R, D$, we have $S \approx \sqrt{D^2 + (2\pi R)^2} = D/\cos\theta$ and, hence, the operator in Eq. (2) can also be expressed as $\hat{H}'(\mathbf{k}_B) = -c^2 \left(\hat{p} - \mathbf{\alpha}\hat{L}\cdot\mathbf{k}_B\right)^2$. The second term inside the parentheses of $\hat{H}'$ represents the contribution from an $l$-dependent synthetic gauge potential induced by the helical structure of the waveguide. Such a synthetic gauge potential induces a coupling between the OAM and the linear momentum. The OAM-$k_B$ interaction lifts the spectral degeneracy of the $\pm q$ vortices, as shown in Fig. 2B for a prototypical system with $a = 3.25$ cm, $R = 32.5$ cm and $D = 65$ cm. The circles denote the full-wave numerical results obtained using the finite-element package COMSOL (*37*). The solid lines denote the analytical results obtained using Eq. (3), which are folded into the first Brillouin zone by using the mapping $k = (D/S)k_B$. The monopole band is not of interest and is thus not shown here. We notice a good match between the analytical and numerical results for the sound vortices $q = \pm 1$. For the straight waveguide, the $q = \pm 1$ vortices are degenerate, whereas this degeneracy is lifted in the helical waveguide. This can also be interpreted as a lateral shift of the bands along opposite directions of $k_B$ due to the OAM-$k_B$ coupling in the presence of a synthetic gauge potential, analogous to the spin-orbit interaction in solid-state materials and cold-atom systems (*38*).

The above phenomenon can also be explained using the geometric phase (that is, Berry phase) introduced to address the adiabatic evolution of states, which has well-known examples in quantum-mechanical systems (*5*). For the helical waveguide with a smooth bending feature, the eigen wave equation can be expressed as

$$-c^2\left[\hat{p}-\alpha\hat{L}\cdot\mathbf{k}(\xi)\right]^2|\mathbf{k}(\xi),l\rangle=-\omega'^2|\mathbf{k}(\xi),l\rangle. \tag{4}$$

Here, we use $|\mathbf{k}(\xi),l\rangle$ to denote the vortex state and $\xi$ is a variable characterizing the position of the vortex inside the waveguide. The above equation is formally analogous to the governing equation of the electron spin evolution under a slowly varying magnetic field, which is expressed as $g\mathbf{s}\cdot\mathbf{B}(t)|\mathbf{B}(t),m_s\rangle=E|\mathbf{B}(t),m_s\rangle$, where $g$ is the gyromagnetic ratio, $\mathbf{s}$ is the electron spin, and $m_s$ is the spin projection along the direction of the magnetic field $\mathbf{B}(t)$. Therefore, one can similarly define a Berry connection (*39*)

$$\mathbf{A}(\mathbf{k})=-i\langle\mathbf{k}(\xi),l|\nabla_\mathbf{k}|\mathbf{k}(\xi),l\rangle=-\frac{l}{k}\cot\theta\hat{\phi}, \tag{5}$$

where $\hat{\phi}$ is the unit direction vector of the azimuthal coordinate $\phi$. The corresponding Berry curvature is simply $\mathbf{F}(\mathbf{k})=\nabla_\mathbf{k}\times\mathbf{A}(\mathbf{k})=l\mathbf{k}/k^3$ (*39*), which represents an effective magnetic monopole at the origin of momentum space, as shown in Fig. 2c. The geometric phase can be obtained as an integral of the Berry connection over the traversed path $C$ in momentum space as shown in Fig. 2C. This gives $\Phi=l\int(1-\cos\theta)d\phi$ in the laboratory frame (*39*) and is proportional to the solid angle subtended by $C$. We note that the geometric phase agrees with Eq. (3), which provides an accumulated phase after time $dt$:

$$\Phi=-\int c\alpha lk dt=l\int(1-\cos\theta)d\phi. \tag{6}$$

Such a geometric phase is essentially induced by the $SO(3)$ variations of vortex propagation direction and hence can be viewed as an acoustic analog of the spin-redirection phase of light with the role of spin played by the OAM of sound vortices.

**Visualizing the spin-redirection geometric phase effect**

To visualize the acoustic spin-redirection geometric phase effect, we performed experiments of sound vortex transport inside a waveguide, as shown in Fig. 3 (see Materials and Methods for details). A sound vortex with $q = +1$ was excited at one end of the waveguide using mini loudspeakers arranged in the same way as in Fig. 1. We measured the output pressure magnitude and phase by a microphone at the other end. Figure 3 (A and D) shows the setup and measured results of the straight cylindrical waveguide. We see that the pressure magnitude has a donut shape with a phase singularity at the center, which is a clear signature of a vortex. Figure 3B shows the same waveguide gently bent on the vertical plane. The resulting pressure field (Fig. 3E) is identical to that of the straight waveguide. This is because the trajectory of the vortex state covers zero area on the sphere in momentum space (Fig. 2C); hence, no geometric phase is induced. When the same waveguide is twisted into a single pitch helix with $\theta = \cos^{-1}(3/4)$, resulting in a 3D configuration as shown in Fig. 3C, the phase profile of the output pressure field (Fig. 3F) shows a rotation compared to the straight and bent case. This is a manifestation of the geometric phase $\Phi = \pi/2$. We note that for a vortex of $q = -1$ with an opposite OAM, the induced geometric phase has an opposite sign.

The observed phenomenon may be viewed as the acoustic analog of circular birefringence induced by the spin-redirection geometric phase. When a linearly polarized dipole mode of the form $e^{i\phi_0} + e^{-i\phi_0}$, where $\phi_0$ denotes the initial polarization direction, is excited at the input of the helical waveguide in Fig. 3C, the right- and left-handed components acquire opposite geometric phase advances, leading to a rotation of the output pressure field by an angle of $\Phi$ as $e^{i\phi_0} + e^{-i\phi_0} \to e^{i\phi_0}e^{i\Phi} + e^{-i\phi_0}e^{-i\Phi} = 2\cos(\phi_0 + \Phi)$. This rotation effect is demonstrated through the experimental and full-wave numerical results shown in Fig. 4. We see that, compared with the input field magnitude, the output field magnitude is rotated approximately by $\pi/2$, in good agreement with the results of Fig. 3C and the analytical results predicted by Eq. 6. The experimental results are also in good agreement with the full-wave simulation results.

**Sound vortex interferometer**

Controlling the transmission of vortex states is useful for applications such as OAM multiplexing, which can improve communication bandwidth (*40, 41*). To demonstrate that the spin-redirection geometric phase can be used to manipulate sound vortex transmissions, we construct an acoustic interferometer as shown in Fig. 5A. The system consists of two waveguides of the same cross-sectional shape. One waveguide is a slightly bent but lies flat on a 2D plane as is defined by the table top. The other waveguide forms a two-pitch helix. The two waveguides meet at two ends and are connected to straight waveguides of the same cross-sectional shape for the input and detection. We excited sound vortices with $q = \pm 1$ at the left input and measured the pressure field at the right output, as shown in Fig. 5A. To characterize the transmission difference between the $q = +1$ and $q = -1$ vortices, we define the parameter $\eta = |t_+ - t_-| / |t_+ + t_-|$, where $t_\pm = \int |p_\pm|^2 dA$ denotes the integral of the square of the pressure field magnitude over the output cross-sectional area for the $q = \pm 1$ vortices. The $t_+$ and $t_-$ are not generally the same because the $q = \pm 1$ vortices acquire different geometric phases after passing through the helical waveguide. The maximum difference between $t_+$ and $t_-$ is achieved when constructive interference occurs in one vortex, whereas destructive interference occurs in the opposite vortex. The measured results are plotted in Fig. 5B as red dots connected by lines, where we observe a marked variation in $\eta$. In Fig. 5C, we show the output pressure profiles for the $q = +1$ (upper panel) and $q = -1$ (lower panel) vortices at the frequency of 3,295 Hz, where $\eta$ reaches a peak. At this point, we see that the two pressure profiles are quite different. For the $q = +1$ input, the output pressure distribution resembles a distorted vortex of the same charge. For the $q = -1$ input, the output is dominated by a monopole field. The emergence of the monopole field is attributed to the symmetry-breaking-induced mode conversion at the branch nodes where the two waveguide branches merge.

For comparison, we performed another measurement in which the same helical waveguide (Fig. 5A) is unwound and lies flat on the same 2D plane defined by the table top. We repeated the measurement with both the $q = +1$ and $q = -1$ input vortices. According to previous analysis, sound vortices acquire no geometric phase propagating through this 2D system. The parameter $\eta$ for this 2D system is plotted in Fig. 5B in blue. In stark contrast to the 3D system with a helical waveguide, it is clearly seen that $\eta$ remains close to zero throughout. This means that the interference spectra are the same for the $q = +1$ and $q = -1$ vortices, as a consequence of the zero geometric phase. In Fig. 5D, we also show the output pressure profiles at 3,295 Hz for the $q = +1$ (upper panel) and $q = -1$ (lower panel) vortices. The two patterns are the mirror images of each other, which is due to the sign of the vortex charge. The output pressure profiles for the peak at approximately $f = 3,230$ Hz (Fig. 5B) have similar features and are shown in Fig. S2 of Supplementary Materials.

It is not easy to precisely parameterize the bending of the waveguide in our experimental system. We performed full-wave simulations for a similar system, and the results are qualitatively the same as the experiment (see Fig. S3 of Supplementary Materials), and lead to the same conclusion.

**DISCUSSION**

The geometric phase effects manifest in classical wave systems in various ways and are defined over different parameter spaces. For example, the manipulation of wave fronts using metasurfaces can be elegantly interpreted as locally induced geometric phases as a result of polarization change induced by subwavelength resonators. The parameter space for these metasurface systems is defined by the Poincaré sphere of polarizations. In photonic or phononic topological insulators, the Brillouin zone serves as the parameter space where a quantized geometric phase can be defined for each discrete band, leading to integral values of topological invariants that can be used to predict the existence and the number of boundary modes. The quantized geometric phase determines the global topological structure of the Bloch wave functions indexed by the Bloch wave vector on the Brillouin torus. Here, the parameter space is defined by the wave vector **k** and the geometric phase is induced by the nontrivial winding of **k**. Such a geometric phase is very general and exists in wave scattering or focusing (*42*). For electromagnetic waves, which are transverse waves, the $\mathbf{k} \cdot \mathbf{E} = 0$ condition guarantees that the polarization direction must accompany the rotation of **k**, and the geometric phase effects arise naturally. Here, we show that sound, as a longitudinal scalar wave, manifests similar geometric phase effects owing to its OAM evolution inside a helical waveguide. For acoustic vortices carrying OAM, we can define a Berry curvature which emerges from an effective magnetic monopole (that depends on the topological charge) in the momentum space with a topological structure that renders the gauge potential globally not well-defined (there is a singularity at the north pole of the momentum sphere). We note that if the bended waveguide lies on a 2D plane, then the trajectory of the vortex states probes zero flux of the gauge field and, hence, their transport behaviors are not affected.

In summary, we theoretically and experimentally demonstrated that airborne sound, as a scalar longitudinal wave without spin, can have spin-redirection geometric phase effects when it travels inside helical waveguides. The spin-redirection geometric phase can be understood as a consequence of non-commutative *SO*(3) rotations or as a result of the adiabatic evolution of the vortex states in the wave vector space. The spin-redirection geometric phase has opposite sign for sound vortices with opposite OAM, leading to different transport properties. Such a geometric phase gives rise to the rotation of the sound pressure field on the transverse plane. The effect can also be considered as an acoustic analog of circular birefringence. Using an interferometer structure, we manipulated the transmission spectrum of sound vortices based on the spin-redirection geometric phase effect. These results demonstrate that, although sound is a scalar wave without spin, the OAM of sound can also have interesting topological properties that may find applications in the manipulations of sound signals.

## MATERIALS AND METHODS

### Wave equation for the helical waveguide

The wave equation for the straight waveguide is $\hat{H}P(\mathbf{r},t) = \partial^2 P(\mathbf{r},t)/\partial t^2$, where $\hat{H} = c^2\nabla^2$. For sound propagating inside a helical waveguide, the wave undergoes an effective rotation of $P' = \hat{R}(\mathbf{n},\phi)P$, where $\hat{R}(\mathbf{n},\phi) = e^{-i\phi\hat{L}\cdot\mathbf{n}}$ is the rotation operator and $\hat{L} = -i\mathbf{r}\times\nabla$ is the OAM operator. The time evolution of the wave state can be expressed as

$$\begin{aligned}\frac{\partial^2 P'}{\partial t^2} &= \frac{\partial}{\partial t}\left(\hat{R}\frac{\partial P}{\partial t} + \frac{\partial \hat{R}}{\partial t}P\right) \\ &= \frac{\partial}{\partial t}\left(\hat{R}\frac{\partial P}{\partial t} - i\hat{L}\cdot\mathbf{\Omega}\hat{R}P\right) \\ &= \left[\hat{R}\frac{\partial^2 P}{\partial t^2} - 2i\hat{L}\cdot\mathbf{\Omega}\hat{R}\frac{\partial P}{\partial t} - \left(\hat{L}\cdot\mathbf{\Omega}\right)^2\hat{R}P\right] \\ &= \left[\hat{R}\hat{H}\hat{R}^{\dagger} + 2\omega\hat{L}\cdot\mathbf{\Omega} - \left(\hat{L}\cdot\mathbf{\Omega}\right)^2\right]P',\end{aligned} \quad (7)$$

where we have used $\hat{H}P = \partial^2 P/\partial t^2$ and $\partial P/\partial t = i\omega P$. Therefore, the operator for the helical waveguide is simply $\hat{H} + 2\omega\hat{L}\cdot\mathbf{\Omega} - (\hat{L}\cdot\mathbf{\Omega})^2$, assuming rotational invariance $\hat{H} = \hat{R}\hat{H}\hat{R}^{\dagger}$. In the laboratory frame, we have $\hat{L}\cdot\mathbf{\Omega} = \mathbf{t}\cdot\hat{L}\Omega(\cos\theta - 1)$, where $\Omega = \partial\phi/\partial t = \sin\theta v_g/R$ with $v_g = \partial\omega/\partial k = c^2k/\omega$ being the group velocity of the wave inside the waveguide. Using the definition of the momentum operator $\hat{p} = -i\nabla$, the final expression for the operator can be rewritten as

$$\hat{H}'(\mathbf{k}) = -c^2\left(\hat{p} - \alpha\hat{L}\cdot\mathbf{k}\right)^2, \quad (8)$$

where $\alpha = |\boldsymbol{\alpha}| = (\cos\theta - 1)\sin\theta c/(R\omega)$ and $\boldsymbol{\alpha}$ has the same direction as $\hat{p}$; and $\mathbf{k} = k\mathbf{t}$ is the longitudinal wave vector of the sound vortices. The second term in the brackets can be understood as a synthetic gauge potential. The corresponding eigen frequency is

$$\omega' = \omega - c\alpha lk. \tag{9}$$

Note that we have assumed adiabatic conditions in the above formulations and Bragg scattering effects are not included.

**Numerical simulations**

The full-wave numerical simulations were performed using the finite element package COMSOL Multiphysics (*37*). All the waveguides had radii of $r = 3.75$ cm in accordance with the experimental setup. For the excitation of the $q = +1$ vortex in Fig. 1 and the simulated rotation effect in Fig. 4, the vortex was excited at one end of the waveguide using four monopole sources arranged on a circle of radius $r = 2.25$ cm. The radiation boundary condition was set at the output end. For the band structure calculations in Fig. 2B, one pitch of the helix was employed with $R = 32.5$ cm and $D = 65$ cm, and periodic boundary condition was applied on both ends.

**Experiments**

The waveguides used in the experiments were polymer tubes, which are slightly flexible so that they can be bent into curved shapes. The wall of the tube was reinforced by steel wire, so that it retained its circular cross-sectional shape when bent. Four mini loudspeakers were connected to four phase-locked signal generators to excite the sound vortices with $q = \pm 1$. To measure the output pressure field, we mounted a miniature microphone on a motorized translational stage to scan the cross section of the waveguide at the output. The radius of the waveguide cross section was $r = 3.75$ cm. The radius and pitch of the helix waveguide in Fig. 3C were $R = 32.5$ cm and $D = 2\pi R\cot\theta$ with $\theta = \cos^{-1}(3/4)$, respectively.

**Acknowledgements:** We thank Prof. Z. Q. Zhang for valuable comments and suggestions.

**Funding:** This work was supported by AoE/P-02/12. S. W. is also supported by a grant from City University of Hong Kong (Project No. 7200549). **Author contributions:** S.W. performed the numerical simulations and analytical calculations. G.M. designed and carried out the experiments. S.W. and G.M. contributed equally to this work. C.T.C. supervised the project. All authors contributed to the discussions and the shaping of the manuscript. **Competing interests:** The authors declare that they have no competing interests. **Data and materials availability:** All data needed to evaluate the conclusions in the paper are present in the paper and/or the Supplementary Materials. Additional data related to this paper may be requested from the authors.

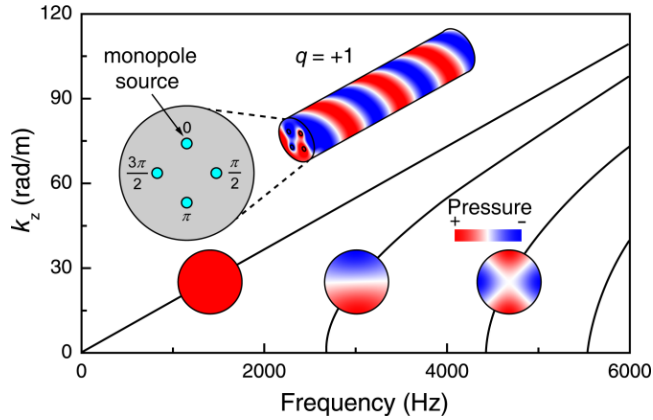

**Fig. 1. Sound vortices inside a circular waveguide.** The solid lines show the dispersion relations of the lowest-order guided modes (monopole, dipole and quadrupole), labeled by the corresponding mode pressure field. A vortex of charge $q = +1$ can be excited using four monopole sources with initial phases $0$, $\pi/2$, $\pi$, $3\pi/2$, and the corresponding pressure field is shown in the inset. The vortex with $q = -1$ can be excited using monopole sources with a reversed order of arrangement.

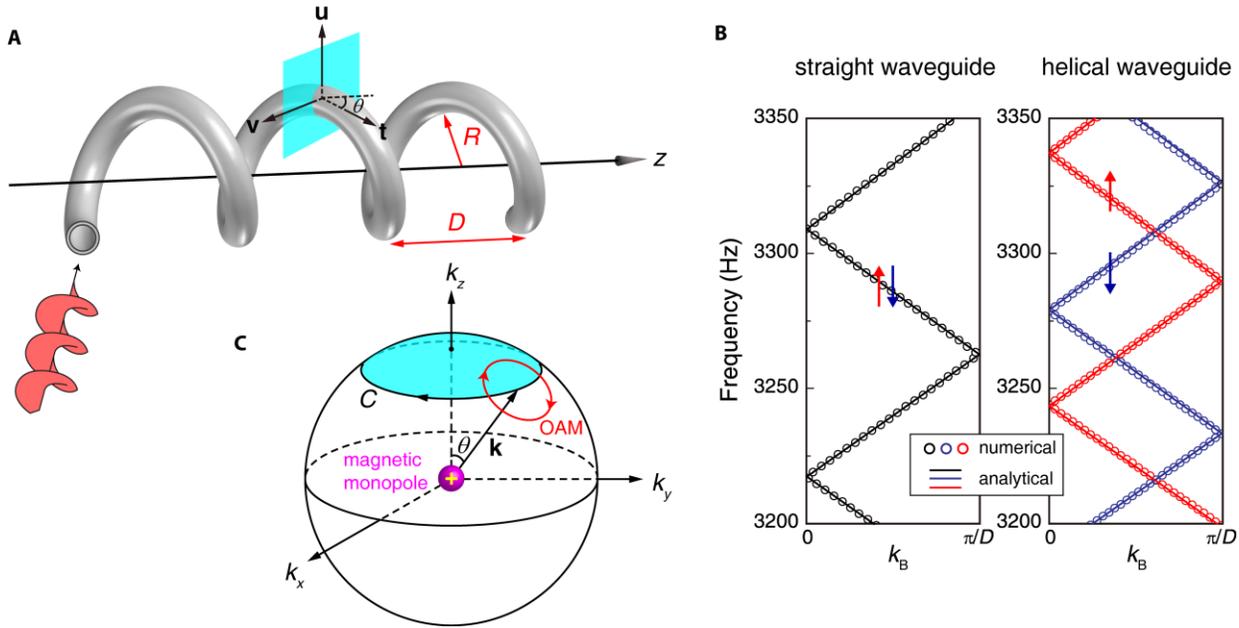

**Fig. 2. Spin-redirection geometric phase induced by helical transport of sound vortices.** (**A**) Sound vortices traveling through a helical waveguide acquire a geometric phase in addition to the normal dynamic phase. Such a phase can be considered as the result of the rotation of the local coordinate system $(\mathbf{u},\mathbf{v},\mathbf{t})$ attached to the waveguide. (**B**) Band structures for a straight waveguide and a helical waveguide showing the degeneracy lifting of the $q=\pm 1$ vortices due to the coupling of the OAM and linear momentum. The monopole band is not shown. (**C**) The degeneracy lifting can be attributed to the spin-redirection geometric phase induced by the evolution of the vortex states in wave vector space. The violet dot at the origin denotes the effective magnetic monopole that depends on the topological charge of the acoustic vortex. The circulation direction of $\mathbf{k}$ depends on the handedness of the helical waveguide. The red circle denotes the circulating direction of the OAM carried by the sound vortex.

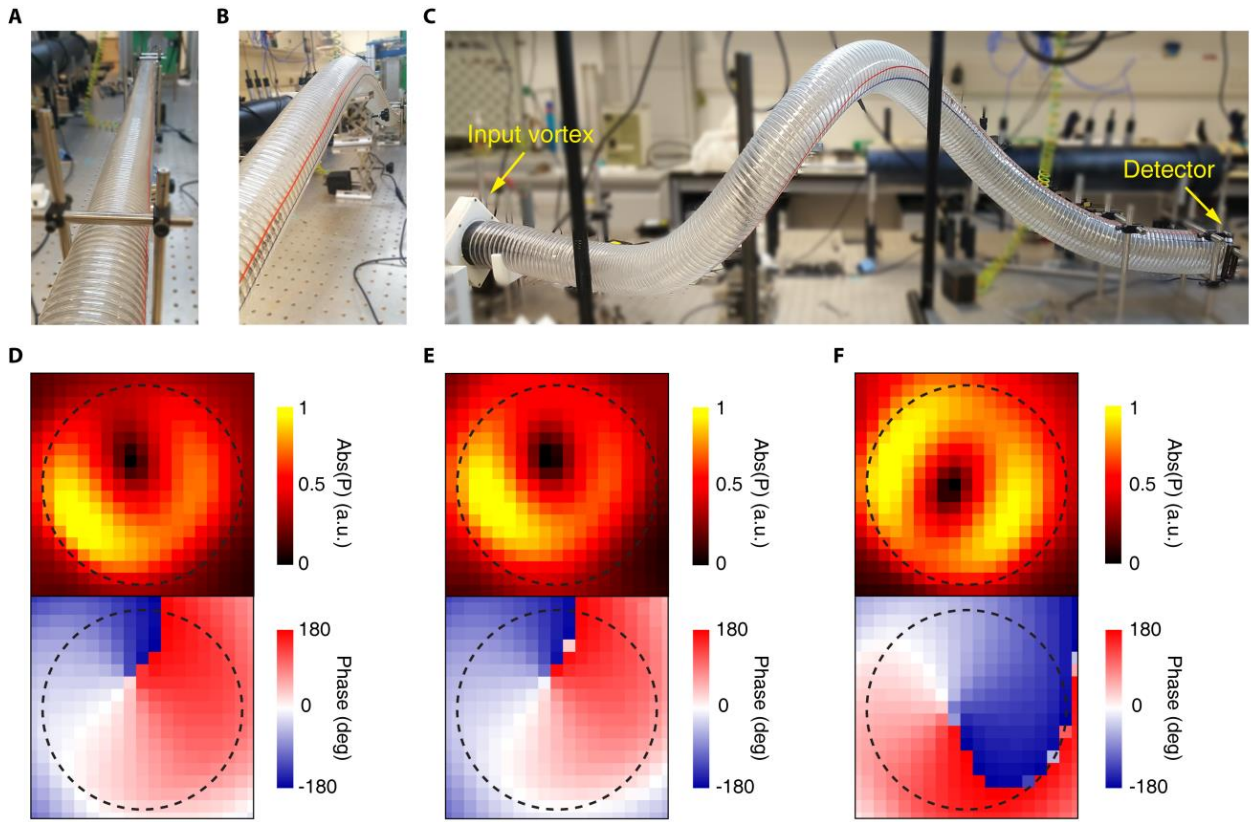

**Fig. 3. Experimental demonstration of the spin-redirection geometric phase for a sound vortex.** (**A** to **C**) Experimental setups for a straight waveguide, a bent waveguide, and a helical waveguide, respectively. (**D** to **F**) Magnitude and phase of the pressure field detected at the end of the waveguide corresponding to (**A**) to (**C**), respectively. The sound vortex is excited using a monopole source array at the input. The dashed circles denote the boundary of the waveguide cross section.

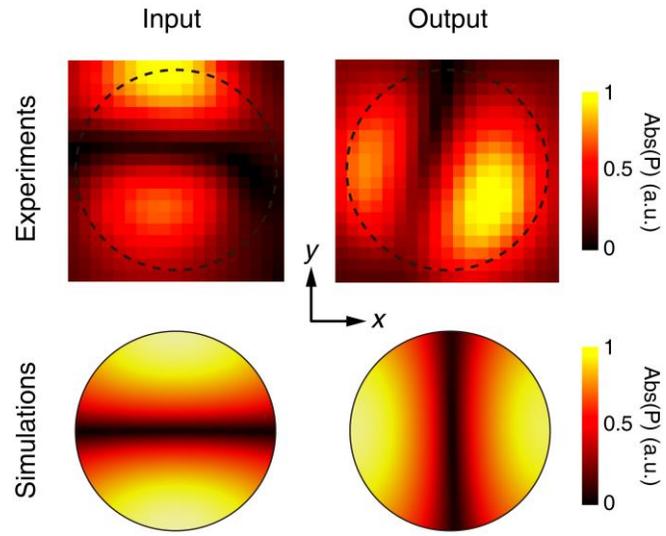

**Fig. 4. Rotation effect induced by the spin-redirection geometric phase.** Comparison between the experimental and full-wave simulation results for the normalized magnitude of the input and output pressure fields when a linear dipole mode along the *y* axis is excited. The system configuration is the same as in Fig. 3C.

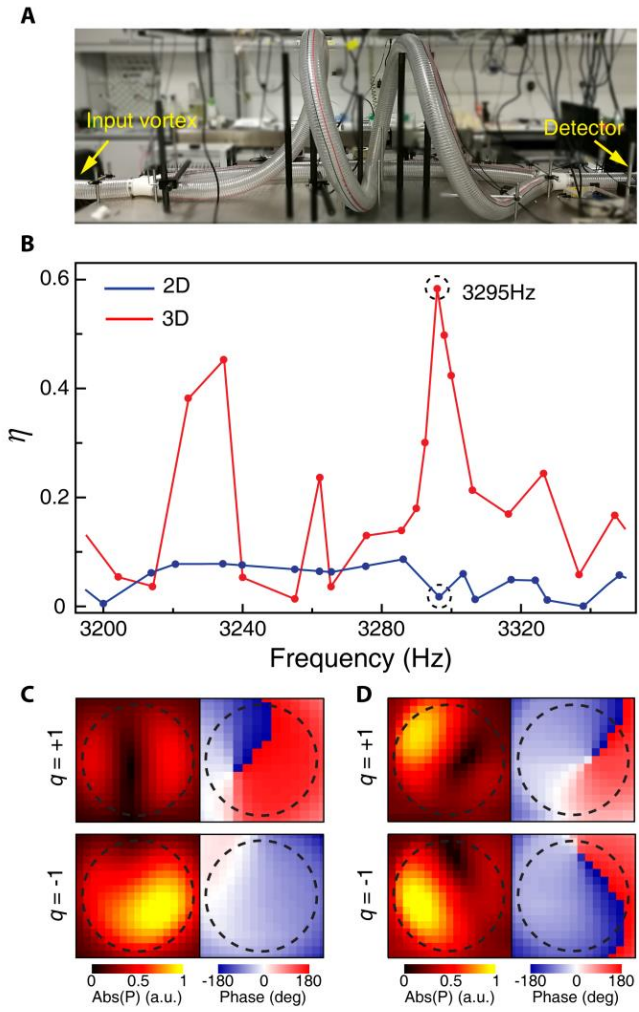

**Fig. 5. Acoustic interferometer based on spin-redirection geometric phases.** (**A**) 3D acoustic interferometer consisting of a helical waveguide and a bent waveguide. (**B**) Transmission contrast $\eta$ (see main text for definition) of the $q = \pm1$ vortices as a function of frequency for the 2D (not shown) and 3D interferometers. The 2D setup is obtained by unwinding the helical waveguide so that the two waveguides lie on a 2D plane. (**C**) Normalized magnitude and phase of the output pressure field of the $q = \pm1$ vortices at the frequency $f = 3,295$ Hz for the 3D case. (**D**) Normalized magnitude and phase of the output pressure field of the $q = \pm1$ vortices at the frequency $f = 3,295$ Hz for the 2D case, which has mirror symmetry.

# Supplementary Materials for

# Topological transport of sound mediated by spin-redirection geometric phase

Shubo Wang[1,2], Guancong Ma[1], and C. T. Chan[1]*

[1]Department of Physics, The Hong Kong University of Science and Technology, Hong Kong, China

[2]Department of Physics, City University of Hong Kong, Hong Kong, China

* Correspondence to C. T. Chan (phchan@ust.hk)

**This PDF file includes:**

- Mapping between waveguide dispersion and band structure.

- Fig. S1. Band structures for helical waveguides with different pitch.

- Fig. S2. An acoustic interferometer based on spin-redirection geometric phases.

- Fig. S3. Full-wave simulations of acoustic interferometers based on spin-redirection geometric phases.

**Mapping between waveguide dispersion and band structure**

To verify our analytic theory, we perform comparisons between the analytical results and full-wave simulation results for the dispersion relations of the vortex modes in the helical waveguide. In the numerical simulations, the band structure is calculated by using one pitch of the helical waveguide and setting periodic boundary conditions on both ends. The analytical results are given in equation (3) of the main text. For comparisons, we fold the analytical dispersion lines back into the first Brillouin zone. This involves a mapping between the guided wave number $k$ and the Bloch wave number $k_B$: $k = (D/S)k_B$, which is given by the condition that the waves undergo the same change of phase when propagating through one pitch of the helical waveguide. Here, $D$ is the pitch and $S$ is the total length of the helical waveguide for one pitch. In the limit of $a \ll R, D$, we have $S \approx \sqrt{D^2 + (2\pi R)^2}$. However, for practical parameters used in the experiments and simulations, this relation is only approximately true. Supplementary Figure 1 shows the comparisons for helical waveguides of different pitch $D$, where we set $a$ = 3.25 cm and $R$ = 32.5 cm. The circles denote full-wave simulation results while the solid lines are analytical results obtained by direct use of $S = \sqrt{D^2 + (2\pi R)^2}$. Such an approximation lead to deviations between the analytical and numerical results, as in the case of $D$ = 65 cm. As $D$ increases, the analytical results gradually approach the numerical results. The black circles denote the band of the monopole mode, which is not of interest.

To obtain accurate values of $S$ for the cases that do not fulfil the condition $a \ll R, D$, we first determine the centre frequency as $\omega = (\omega_+ + \omega_-)/2$, where $\omega_\pm$ denote the numerical results for the frequency of the $\pm q$ mode. This essentially gives the band structure of the corresponding straight waveguide. By fitting the analytical expression for the dispersion relation of the straight waveguide, i.e., $\omega = c\sqrt{k_r^2 + k^2}$, to this centre frequency band structure, we obtain the correct values of $S$, which can be used for the mapping of $k = (D/S)k_B$. The results in Fig. 2b of the main text are obtained in this way.

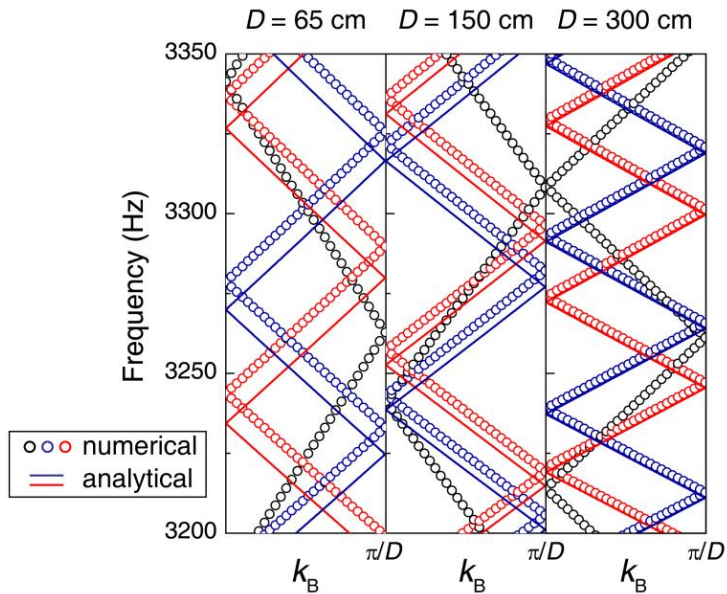

**Fig. S1. Band structures for helical waveguides with different pitch.** The circles denote the full-wave simulation results while the solid lines denote the analytical results, which are obtained by folding the band dispersion of equation (3) into the first Brillouin zone using the mapping of $k = (D/S)k_B$. Here, we make approximations and set $S = \sqrt{D^2 + (2\pi R)^2}$.

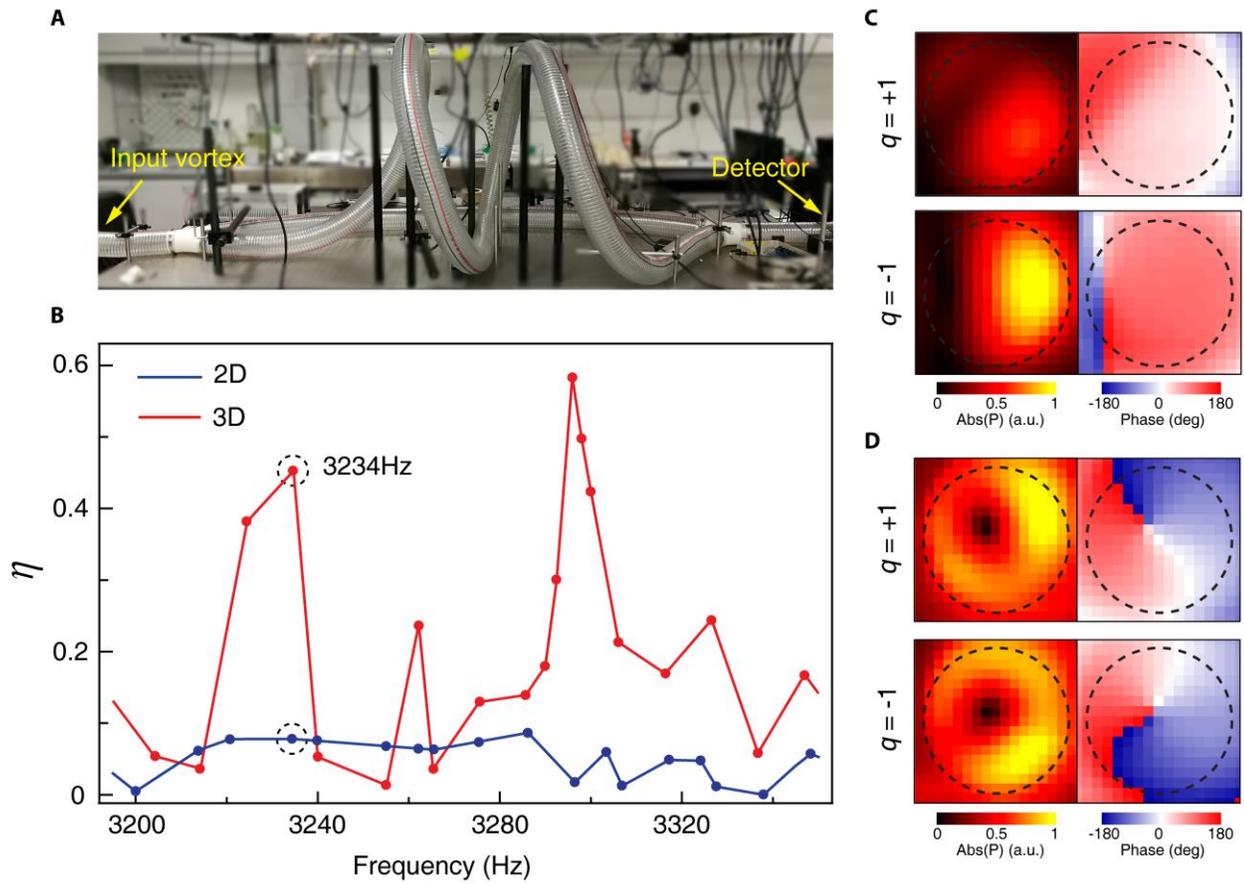

**Fig. S2. An acoustic interferometer based on spin-redirection geometric phases.** (**A**) A 3D acoustic interferometer consisting of a helical waveguide and a bent waveguide. (**B**) The transmission contrast $\eta$ (see main text for definition) of the $q = \pm1$ vortices as a function of frequency for the 2D (not shown) and 3D interferometers. The 2D setup is obtained by unwinding the helical waveguide so that the two waveguide branches lie on a 2D plane. (**C**) The normalized magnitude and the phase of the output pressure field of the $q = \pm1$ vortices at the frequency $f = 3{,}234$ Hz for the 3D case. (**D**) The normalized magnitude and phase of the output pressure field of the $q = \pm1$ vortices at the frequency $f = 3{,}234$ Hz for the 2D case, which has mirror symmetry.

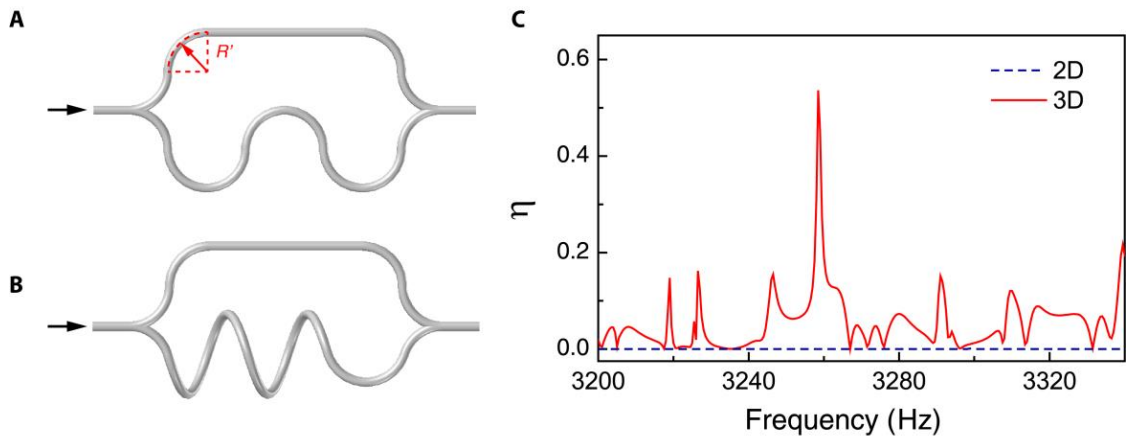

**Fig. S3. Full-wave simulations of acoustic interferometers based on spin-redirection geometric phases.** The configurations of the 2D and 3D interferometers are shown in (**A**) and (**B**), respectively, where the $q = \pm 1$ vortices are excited at the left end. The arc parts of the waveguide in both (**A**) and (**B**) have radii of $R' = 34$ mm. The helical part has radius $R$ and pitch $D$ with the same values as in the experiments. (**C**) The simulation results of the transmission contrast $\eta$ for the 2D and 3D interferometers.